\documentclass[12pt]{article}
\usepackage{amssymb}

\newcommand{\be}{\begin{equation}}
\newcommand{\ee}{\end{equation}}
\def\bea{\begin{eqnarray}}
\def\eea{\end{eqnarray}}
\newcommand{\bn}{\begin{eqnarray}}
\newcommand{\en}{\end{eqnarray}}
\newcommand{\nn}{\nonumber}
\newcommand{\no}{\noindent}

\newcommand{\p}{\partial}
\def\bea{\begin{eqnarray}}
\def\eea{\end{eqnarray}}
\newcommand{\beq}{\begin{eqnarray}}
\newcommand{\eeq}{\end{eqnarray}}

\begin{document}

\title{\textbf{A note on higher rank descriptions of massless and massive spin-1 particles}}
\author{{ D. Dalmazi\footnote{denis.dalmazi@unesp.br} $\,\,$,  F.A. da Silva Barbosa\footnote{felipe-augusto.barbosa@unesp.br} and A. L. R. dos Santos$^{2}$\footnote{alessandroribeiros@yahoo.com.br}}
\\
\textit{{ UNESP - Campus de Guaratinguet\'a - DFI} }\\
\textit{{CEP 12516-410, Guaratinguet\'a - SP - Brazil.} }\\}
\date{\today}
\maketitle

\begin{abstract}

The Maxwell theory can be written as a first order model with the help of a two-form auxiliary field, such master action allows the proof of duality between $1$-form and $D-3$ forms. Here we show that the replacement of the two-form auxiliary field by an arbitrary (non symmetric) rank-2 tensor leads to a new massless spin-1  dual  theory in terms of a partially antisymmetric  rank-3 tensor. In the massive spin-1 case we have a non symmetric generalization of the massive two-form theory (Kalb-Ramond). The coupling of the massive non symmetric spin-1 model to matter fields is investigated via master actions. 

We also show that massive models with severe discontinuity in their massless limit can also be obtained from Kaluza-Klein dimensional reduction of massless higher rank tensors which become Stueckelberg fields after the reduction.

\end{abstract}

\newpage

\section{Introduction}

Independently of the phenomenological viability of modern theories containing  massive gravitons \cite{drgt,hs3}, see the review work \cite{lh} and the recent work \cite{hm}, the existence of a possible graviton mass should be investigated as a matter of principles. It turns out that even as a free theory, massive spin-2 particles are non trivial. In order to have  the correct number of degrees of freedom $(2\, s +1=5) $ and a stable classical field theory, the mass term must be fine tuned \cite{fp}  in a unique way \cite{pvn} if we make use of a symmetric rank-2 tensor\footnote{The use of a more general non symmetric rank-2 tensor allows different mass terms \cite{nfp,rank2}.}. The interpretation of the metric fluctuation about a flat background, at linearized level, as a massive spin-2 particle leads to further difficulties since it never fits the experimental data even if we had a tiny mass \cite{vdv,zak}. The cure comes from non linear terms via the Vainshtein \cite{vain} screening mechanism which on its turn introduces in general a ghost mode \cite{db}. 

The spin-2 self-interaction (graviton potential) can be however, judiciously chosen \cite{drgt} in order to avoid ghosts \cite{drgt,hr}. This choice is behind the more viable bimetric model of \cite{hrbm}. It is extremely hard to introduce self-interacting terms which do not turn on the field components that must remain without dynamics in order that the correct number of degrees of freedom (five) of a massive spin-2 particle is achieved.
An underlying key ingredient behind the graviton potential is the Galileon symmetry \cite{nrt}. The Galileon self-interacting Lagrangians for scalar fields lead to second order field equations for the helicity zero mode of the graviton thus, avoiding the Ostrogadsky instability and allowing the expected 5 degrees of freedom. As a general question one might think of going beyond scalar fields and search for Galileon-like self-interacting Lagrangians with higher rank fields still leading to second order field equations with the correct number of degrees of freedom. In the case of massless p-forms it has been shown \cite{pforms} that no such generalization does exist in $D=4$. In particular, for massless spin-1 particles   described by a vector field (1-form)
there is a no-go theorem \cite{dgmw}  for Galileon Lagrangians in arbitrary dimensions. However, for massive particles, already in the 1-form case  one has been able to construct \cite{gpt,gph} generalized Proca (GP) models in $D=4$ with the help of Galileon-like vertices.  The GP models have derivative self-interactions but second order field equations and the time component of the vector field $A_0$ remains non dynamic, thus  warranting   $2s+1=3$ propagating degrees of freedom. Those models, when coupled to gravity, have very interesting cosmological and astrophysical consequences. The massive vector field may be a viable dark energy candidate, see e.g. \cite{fhkmt,jhknt,sky}. It may relieve the Hubble constant tension \cite{h01,h02} as shown in \cite{sky}. There are also phenomenological applications in black hole physics \cite{bh1,bh2}, see \cite{shz} regarding stability problems.
Further generalizations of a self-interacting Proca model \cite{gpn} and of multiple Proca ($m\ne 0$) and Maxwell ($m=0$) vector fields interacting with each other have been suggested \cite{mayprd20}.

Massive spin-1 particles can also be described in terms of a two-form (antisymmetric rank-2 tensor) \cite{kr} which is by the way dual to topologically massive BF models in $D=3+1$, as shown in \cite{ak98,hs}. The BF theory finds applications also in condensed matter physics as in the case of topological ordering \cite{hos,cpp} and in topological insulators \cite{blasi}. 

More recently one has also found a description of massive spin-1 particles in terms of a symmetric rank-2 tensor \cite{renato1}. Since the massless limit of such models differs from the massless limit of the Proca model and the building of self-interaction vertices strongly depends on the tensor structure of the fundamental field it is interesting to search for Galileon-like Lagrangians for those higher rank fields. Indeed, this point has been addressed in \cite{2forms} for antisymmetric fields and   some Galileon-like Lagrangians for massive 2-forms (Kalb-Ramond fields)  have been found  in $D=4$. They stop at second order in the field derivatives, differently from the vector case which goes through the third order \cite{gpt,gph}. Higher order terms become total derivatives. We firmly believe that this difficult relies on the specific antisymmetric tensor structure. In the present work  we find a description of massive spin-1 particles via a non symmetric rank-2 tensor which generalizes previous descriptions \cite{kr,renato1}  and may be used in the search for more general Galileon-like terms for massive spin-1 models. The non symmetric case can not be reduced to the previous ones and the lack of symmetries in tensor indices may allow us to build up higher order (in derivatives) self-interacting vertices which may lead to different cosmological and astrophysical consequences after coupling to gravity. We also find a higher rank description of massless spin-1 particles.

In section 2, starting from a massless master action including a non symmetric tensor and the usual vector field we derive   a new description of a massless spin-1 particle dynamics in terms of a second order theory described by a rank-3 partially antisymmetric tensor. We connect gauge invariants of such theory with the usual electric and magnetic fields via dual maps. In section 3 we add a  mass term and obtain a new description of massive spin-1 particles in terms of a non symmetric rank-2 tensor. We add interactions with matter fields in the master action approach and find dual maps between the usual Proca theory and the new non symmetric model, both linearly coupled to matter currents. In section 4 we comment on the discontinuity of the massless limit and show how massive theories with rather severe massless discontinuities like spin-jumping and abrupt change of particle content can still be derived from dimensional reductions of higher rank massless models with the same spin. In section 5 we draw our conclusions and comment on the coupling to gravity. In the appendix we make a detailed canonical analysis of the new massive non symmetric model (\ref{lns}) and show its equivalence to the Proca theory. 

\section{Massless spin-1 models}

We start from a rather general first order version of the Maxwell theory in $D$ dimensions where the auxiliary field is a general (nonsymmetric) rank-2 tensor 
$e_{\mu\nu}$. The constant coefficients $(a,b,c,d,f,g)$ must be such that the Gaussian integrals over $e_{\mu\nu}$ lead to the Maxwell theory,

\be S = \int d^D x \left[ \left( f\, e_{\mu \nu} + g\, e_{\nu \mu} \right) \p^\mu A^\nu +  a\, e\, \partial^\mu A_\mu  + b\, e_{\mu \nu} e^{\mu \nu} + c\, e_{\mu \nu} e^{\nu \mu} + d\, e^2  \right], \label{S1} \ee

\no where $e=\eta^{\mu\nu} e_{\mu\nu}$. Without loss of generality, we can always set $a=0$ by an invertible field redefinition  $e_{\mu\nu} \to e_{\mu\nu}+ \lambda\, \eta_{\mu\nu} \, e $ with an appropriate choice of $\lambda$ such that $\lambda\ne -1/D$. Next, we could rotate the rank-2 tensor such that $f\, e_{\mu \nu} + g\, e_{\nu \mu} \to \tilde{e}_{\mu\nu}$, however this is only invertible if $f\ne \mp g$. 
The first case $f=-g$, which naturally leads to $b=-c$ and $d=0$, corresponds to an antisymmetric (two-form) auxiliary field\footnote{Throughout this work we use
$\eta_{\mu\nu}=(-,+,+,\cdots,+)$,
$e_{(\alpha\beta)}=(e_{\alpha\beta}+e_{\beta\alpha})/2$ and
$e_{[\alpha\beta]}=(e_{\alpha\beta}-e_{\beta\alpha})/2$.} $B_{\mu\nu}\equiv e_{[\mu\nu]}=-B_{\nu\mu}$. Choosing $g=-2\,c=1/4$ we have,

\be {\cal L}_{B} = \frac{1}{4}B^{\mu\nu}B_{\mu\nu}-\frac{1}{2}B^{\mu\nu}(\partial_{\mu}A_{\nu}-\partial_{\nu}A_{\mu}) \label{lb} \ee

\no Integrating over $B_{\mu\nu}$ we obtain the usual formulation of the Maxwell theory ${\cal L}_{max}(A)=-F_{\mu\nu}^2(A)/4$ while the integral over $A_{\mu}$ leads to the constraint  $\partial_{\mu}B^{\mu\nu}=0$ whose general solution is $B^{\mu\nu}=\partial_{\alpha}T^{[\alpha\mu\nu]}$ where $T^{[\alpha\mu\nu]}$ is a completely antisymmetric but otherwise arbitrary tensor. Back in (\ref{lb}) we obtain the dual Lagrangian density 

\be {\cal L}_{TT} =  \frac{1}{4} (\partial_{\alpha}T^{[\alpha\mu\nu]})^2 \label{ltt} \ee

\no Since we can always write $T_{[\alpha\mu\nu]}=\epsilon_{\alpha\mu\nu\beta_1\beta_2 \cdots \beta_{D-3}}\tilde{T}^{\beta_1\beta_2 \cdots \beta_{D-3}}$, the duality between Maxwell and ${\cal L}_{TT}$ corresponds to the known  duality, see e.g. \cite{hl} and \cite{renato1}, between massless $1$-forms and $(D-3)$-forms. In particular, in $D=4$ we go back to the usual Maxwell vector theory (self-duality).

The second case where $f=g$, altogether with $b=c$ and $d=-2\,c/(D-1)$, is much less known, see \cite{kmu}. Now the auxiliary field is a symmetric tensor $W_{\mu\nu}\equiv e_{(\mu\nu)} $, with $g=2\,c=1$ we have,

\be {\cal L}_W =  W^{\mu \nu} W_{\mu \nu} - \frac{W^2}{D-1} + 2 \, W^{\mu \nu} \partial_{( \mu} A_{ \nu )} \label{lw} \ee

\no Integrating over $W_{\mu\nu}$ we have the Maxwell theory while integrating over $A_{\mu}$ we have the constraint $\p^{\mu}W_{\mu\nu}=0$ whose solution \cite{dt} may be written in terms of a rank-4 tensor with the same index symmetries of the Riemann curvature tensor, namelly $W_{\mu\nu} = \partial^\alpha \partial^\beta B_{[\mu \alpha] [\nu \beta]}$. Back in (\ref{lw}) we obtain a $D$-dimensional generalization of the $D=4$ massless spin-1 model of \cite{dts},

 \be {\cal L}_{DTS} =  \left(\partial^\alpha \partial^\beta B_{[\mu \alpha] [\nu \beta]} \right)^2 - \frac{ \left( \partial^\alpha \partial^\beta B^{[\mu}_{\quad\!\!\alpha ] [\mu \beta ]} \right)^2}{D-1} \label{ldts} \ee

\no In $D=2+1$ we can rewrite ${\cal L}_{DTS}$ in terms of a symmetric rank-2 tensor via $B^{[\mu\alpha][\nu\beta]}= \epsilon^{\mu\alpha\lambda}\epsilon^{\nu\beta\sigma}h_{\lambda\sigma}$ such that ${\cal L}_{DTS}$ is proportional to  the linearized version   ($g_{\mu\nu} = \eta_{\mu\nu}+h_{\mu\nu}$) of the fourth order  term of the ``New Massive Gravity'' of \cite{bht}, i.e., ${\cal L}_{DTS} \sim \left( R_{\mu\nu}^2 - 3\, R^2/8\right)_{hh}$ which is known to be dual to the Maxwell theory \cite{bht2}. In $D=3+1$ the equivalence of (\ref{ldts}) to the Maxwell theory is shown in \cite{dts}. 

Now we turn to the non symmetric (NS)  case $f\ne \pm g$ which,  to the best we know, has not appeared before in the literature. Without loss of generality we assume henceforth $b(b^2-c^2)\ne 0$ with  $(a,d) =(0,(b+c)/(1-D))$ and $(f,g)=(1,0)$,

\be S_{NS} = \int d^D x \left\lbrack e_{\mu \nu}  \partial^\mu A^\nu + b\, e_{\mu \nu} e^{\mu \nu} + c\, e_{\mu \nu} e^{\nu \mu} - \frac{(b+c)}{D-1} e^2  \right\rbrack \label{sns}\ee

\no The action $S_{NS}$ is invariant under $U(1)$ symmetry:

\be \delta A_{\mu} = \p_{\mu}\Lambda  \quad ; \quad \delta e_{\mu\nu} = \frac{\left(\eta_{\mu\nu}\Box - \p_{\mu}\p_{\nu} \right) \Lambda}{[2(b+c)]} \equiv \frac{\Box\theta_{\mu\nu}\Lambda}{2(b+c)} \quad . \label{dL} \ee

\no After Gaussian integrating over $e_{\mu\nu}$, supposing $b/(b^2-c^2) >0$,  we obtain the Maxwell theory with the correct overall sign,

\be {\cal L}_{max}(b,c) = -\frac b{8(b^2-c^2)} F_{\mu\nu}^2 (A) \quad . \label{lmaxbc} \ee

\no At this point it is interesting to remind the reader that (\ref{lb}), (\ref{lw}) and (\ref{sns}) have their correspondence in the linearized spin-2 ( linearized gravity) case. It is known that general relativity can also be formulated {\it a la } Einstein-Cartan   as a first order theory in terms of the spin connection and vierbein, at linearized level we have ${\cal L}_{EC}\sim \omega \cdot \omega + \omega \, \p e $ with $\omega_{\mu \, a\, b}= -\omega_{\mu \, b\, a}$  and also in terms of the Christoffel symbol and the metric  {\it a la} Palatini ${\cal L}_{P}\sim \Gamma \cdot \Gamma + \Gamma \, \p g $ with, in the torsionless case, $\Gamma^{\mu}_{\alpha\beta} = \Gamma^{\mu}_{\beta\alpha}$.  At linearized level those cases seem to work like spin-2 analogues of (\ref{lb}) and (\ref{lw}) respectively while the inclusion of torsion, see the review works \cite{olmo1,olmo2}, is the analogue of (\ref{sns}) as if we had just suppressed the $\mu $ index from the spin connection and from the Christoffel symbol.

 If we integrate the vector field $A_{\mu}$ in (\ref{sns}) we generate the constraint $\p^{\mu}e_{\mu\nu} = 0 $ whose solution can be written in terms of a partially antisymmetric rank-3 tensor, $B_{[\alpha\mu]\nu}=- B_{[\mu\alpha]\nu}$, i.e., 

\be e_{\mu\nu}[B] \equiv \p^{\alpha}B_{[\alpha\mu]\nu} \label{eb}\ee 

\no Plugging back in $S_{NS}$ we have the new massless spin-1 dual model

\bea {\cal L}_{BB} &=& b\, \partial^\beta B_{[\beta \mu] \nu} \partial_{\alpha} B^{[\alpha \mu] \nu} + c\, \partial^\beta B_{[\beta \mu]\nu} \partial_{\alpha} B^{[\alpha \nu] \mu} - \frac{(b+c)}{D-1} \left(\partial^\gamma B_\gamma \right)^2 \label{lbb} \\
&=& b\, e_{\mu \nu}[B] e^{\mu \nu}[B] + c\, e_{\mu \nu}[B] e^{\nu \mu}[B] - \frac{(b+c)}{D-1} e^2[B] \quad , \label{lbb2}
 \eea 

\no where $B_{\gamma}=\eta^{\mu\nu}B_{[\gamma \mu] \nu}$. Henceforth we keep the pair $(b,c)$ arbitrary, except for the condition $b(b^2-c^2)> 0$, since it can not be changed by any local field redefinition. 

The second order equations of motion $\delta S_{NS}/\delta B_{[\gamma\mu]\nu} =0$ can be rewritten as a zero curvature condition,

\be
    \partial_\gamma \bar{e}_{\mu \nu}[B] - \partial_\mu 
    \bar{e}_{\gamma \nu} [B]=0, \label{eombar}
\ee
where 
\be
    \bar{e}_{\mu \nu} [B]\equiv b\, e_{\mu \nu}[B] + c\, e_{\nu \mu}[B] - \frac{(b+c)}{D-1}\, e[B] \, \eta_{\mu \nu} \label{ebar}.
\ee

\no The general solution of (\ref{eombar}) is $\bar{e}_{\mu \nu} = \partial_{\mu} A_{\nu}$ for some vector field $A_{\mu}$. Back in (\ref{ebar}) we have 

\be
e_{\mu \nu} =  \frac{1}{b^2-c^2} \left( b \, \partial_\mu A_\nu - c \,\partial_\nu A_\mu \right) - \frac{\eta_{\mu \nu}}{b+c} \partial^\mu A_\mu, \label{emna}
\ee

\no The Maxwell equations now follow from an identity,

\be \partial^\mu e_{\mu \nu}[B] = \frac{b}{b^2-c^2} \p^{\mu}F_{\mu\nu}(A) = \p^{\mu} \partial^\alpha B_{[\alpha \mu ] \nu} = 0. \label{id}
\ee

 Before we proceed, a remark is in order. Namely, since the equations of motion (\ref{eombar}) are invariant under $\delta_C \bar{e}_{\mu\nu}[B] = \p_{\mu} C_{\nu} $  for any vector $C_{\mu}$, one might think that the solution $\bar{e}_{\mu\nu}=\p_{\mu}A_{\nu}$ is pure gauge. This is however, not the case since there is no local transformation of the fundamental field $\delta_C B_{[\alpha \mu ] \nu} $ which might lead to $\delta_C \bar{e}_{\mu\nu}[B] = \p_{\mu} C_{\nu} $ in general. This is only possible if $C_{\nu} = \p_{\nu}\Lambda $ which represents the $U(1)$ symmetry of (\ref{lbb}). The action $S_{BB}$ corresponding to (\ref{lbb}) is invariant under 
the gauge transformations:

\be  \delta B_{[\beta \mu] \nu} = \left[\eta_{\beta \nu } \partial_\mu \Lambda - \eta_{\mu \nu} \partial_\beta \Lambda \right] + \Omega_{[\beta \mu] \nu}^{T} \quad , \label{dB} \ee

\no where $\p^{\beta}\Omega_{[\beta \mu] \nu}^{T}=0$. Notice that the previous remark also points to the fact that there is no local formula for the fundamental field $B_{[\alpha \mu] \nu}$ in terms of the photon field $A_{\mu}$. This is also true for the higher rank model ${\cal L}_{DTS}$ given in (\ref{ldts}). 

Now, in order to compare correlation functions of gauge invariants of the usual Maxwell theory  with the corresponding ones of the new model  ${\cal L}_{BB}$ we add sources to (\ref{sns}). This is a bit subtle due to the higher rank and the gauge symmetry. First, from (\ref{dL}) we see that the anti symmetric components $e_{[\mu\nu]}[B]$ are gauge invariants. So one can formally define a gauge invariant generating function starting with
the Lagrangian density below in the path integral, 

\be {\cal L}_{NS}[J] =b\, e_{\mu \nu} e^{\mu \nu} + c\, e_{\mu \nu} e^{\nu \mu} - \frac{(b+c)}{D-1} e^2 + e_{\mu \nu}  \partial^\mu A^\nu  + e_{[\mu\nu]}J^{[\mu\nu]} \label{lj} \ee

\no where  $J^{[\mu\nu]}$ is an arbitrary anti symmetric external source. If we first integrate over $e_{\mu\nu}$ in the path integral we obtain the effective action:

\be {\cal L}^{EF}_{Max}[J_{[\mu\nu]}] = -\frac{b F_{\mu \nu} F^{\mu \nu} }{8(b^2-c^2)} - \frac{ J^{[\mu \nu]} F_{\mu \nu} }{4 (b-c)} - \frac{J_{[\mu \nu]} J^{[\mu \nu]} }{4(b-c) }  \quad . \label{lmaxj} \ee

\no On the other hand, if we first integrate on the vector field $A_{\mu}$ and solve the functional constraint $\p^{\mu}e_{\mu\nu}=0$  we obtain 

\be {\cal L}^{EF}_{BB}[J] = b \,\partial^\beta B_{[\beta \mu] \nu} \partial_{\alpha} B^{[\alpha \mu] \nu} + c \,\partial^\beta B_{[\beta \mu]\nu} \partial_{\alpha} B^{[\alpha \nu] \mu} - \frac{(b+c)}{D-1} \left(\partial^\gamma B_\gamma \right)^2  +  \partial^\beta B_{[\beta \mu ] \nu} J^{[\mu \nu]} \quad . \label{lbbj} \ee

\no From functional derivatives of the effective generating functions obtained from (\ref{lmaxj}) and (\ref{lbbj}) with respect to the sources at $J_{[\mu\nu]}=0$ we obtain the correspondence between correlation functions of gauge invariants:

\be \left(\frac{-1}{4(b-c)}\right)^N  \left\langle F_{\mu_1\nu_1}(x_1) \cdots F_{\mu_N\nu_N}(x_N)\right\rangle_{Max} = \left\langle e_{[\mu_1\nu_1]}[B(x_1)] \cdots e_{[\mu_N\nu_N]}[B(x_N)]\right\rangle_{BB} + \quad c.t. \label{ct} \ee

\no where $c.t.$ stands for contact terms.  The previous gauge invariant result leads to the following  dual map between the Maxwell theory and the ${\cal L}_{BB}$ model,

\be  -\frac{F_{\mu \nu}(A)}{4(b-c)} \longleftrightarrow \frac{1}{2} \left( \partial^\alpha B_{[\alpha \mu] \nu} - \partial^\alpha B_{[\alpha \nu] \mu} \right) = e_{[\mu\nu]}[B] \label{map1}\ee

\no Notice that the classical solution of the equations of motions given 
in (\ref{emna}) leads to $e_{[\mu\nu]}=F_{\mu\nu}/[2(b-c)]$. So the vector field found by solving (\ref{eombar}) coincides with the vector field of the dual Maxwell theory up to an irrelevant overal  factor $-1/2$. If we had written the solution of the equations of motion as $\overline{e}_{\mu\nu} = -\p_{\mu}A_{\nu}/2$ there would be full agreement.

 Although we know that all local $U(1)$ invariants of the Maxwell theory are encoded in the field strength $F_{\mu\nu}(A)$, it turns out that not all gauge invariants of the $S_{BB}$ theory are represented by $e_{[\mu\nu]}[B]$. From the gauge transformations (\ref{dB}) we have 
$\delta_{\Lambda} \, e_{\mu\nu}[B]=\p_{\nu}\p_{\mu}\Lambda - \eta_{\mu\nu} \Box \Lambda $. After we eliminate the gauge parameter via $\Lambda = \delta_{\Lambda } \, e_{00}[\Lambda ]/\nabla^2 $ and plug it back in the $D(D+1)/2$ equations  $\delta_{\Lambda} \, e_{(\mu\nu)}[B]$ stemming from (\ref{dB}) we end up with $D(D+1)/2 -1$ gauge invariants built out of the symmetric components $e_{(\mu\nu )}[B]$, i.e., 

\bea I_j &=& \nabla^2 e_{(0j)}[B] - \p_0\p_j e_{00}[B] \quad , \label{ij} \\
I_{ij} &=& \nabla^2 e_{(ij)}[B] + (\Box \delta_{ij} - \p_i\p_j )e_{00}[B] \label{iij} \eea 

\no The reader can check their invariance under (\ref{dB}). Those extra gauge invariants raise the question about the existence of extra physical quantities in the $S_{BB}$ model which may not be mapped into functions of the field strength of the Maxwell theory. In order to investigate that issue we have replaced the anti symmetric source term in (\ref{lj}) by a symmetric one:
 $e_{(\mu\nu)}J^{(\mu\nu)}$.  Due to the $U(1)$ gauge symmetry (\ref{dL}) the newly introduced symmetric source must satisfy the scalar constraint 
 
 \be \Box \theta^{\mu\nu}J_{(\mu\nu )} =0 \quad . \label{thetamn} \ee
 
 \no If we integrate over $e_{\mu\nu}$ in the path integral we obtain the effective action, compare with (\ref{lmaxj}),
 
 \be {\cal L}^{EF}_{Max}[J_{(\mu \nu)}] = -\frac{b F_{\mu \nu} F^{\mu \nu} }{8(b^2-c^2)} + \frac{ J^{(\mu \nu)} (\eta_{\mu \nu}\p \cdot A - \p_{\mu}A_{\nu}) }{2 (b+c)} + \frac{J^2 - J_{(\mu \nu)}^2 }{4(b+c) }  \quad . \label{lmaxj2} \ee
 
 \no On the other hand, integrating $A_{\mu}$ we deduce the effective action (\ref{lbbj}) with the replacement $J_{[\mu \nu]} \to J_{(\mu \nu)}$. Both effective actions are $U(1)$ invariant due to the constraint (\ref{thetamn}) which allows the elimination of one of the $D(D+1)/2$ components of $J_{(\mu\nu)}$. The remaining independent components couple precisely to the invariants\footnote{This is similar to the usual Maxwell theory where the decomposition $(J_0,J_k)=(\rho,S_k+\p_k S)$, with $\p_kS_k=0$, and the constraint $\p_{\mu}J^{\mu}=0$ allow us to eliminate $S=\dot{\rho}/\nabla^2$ and write $J^{\mu}A_{\mu}= S_k\p_iF_{ik}/\nabla^2 + \rho \p_kF_{0i}/\nabla^2$.} (\ref{ij}) and (\ref{iij}). Functional derivatives with respect to the unconstrained components of $J_{(\mu\nu)}$ lead to the following dual maps which hold up to contact terms:
 
 \bea \frac{\p_i (\p_j F_{0i}-\p_0 F_{ij})}{2(b+c)} &&\longleftrightarrow
 I_j \quad , \label{map2} \\
 \frac{\p_k (2\, \delta_{ij}\p_0 F_{0k}-\p_i F_{kj} -\p_j F_{ki})}{2(b+c)} &&\longleftrightarrow
 I_{ij} \quad , \label{map3}  \eea
 
 In summary, all gauge invariants of the ${\cal L}_{BB}$ model built out of $e_{\mu\nu}[B]$  can be written in terms of electric and magnetic fields and their derivatives. There are no new gauge invariants. So the higher rank free model  ${\cal L}_{BB}$ given in (\ref{lbb}) is equivalent to the usual vector Maxwell theory. 
 It is of course not clear if such equivalence at free level holds also after the introduction of self-interacting vertices or how would we map vector vertices into vertices of the rank-3 tensor $B_{[\mu\nu]\alpha}$. In particular, we should revisit the no-go theorem of \cite{dgmw} for massless vector Galileons replacing the basic gauge invariant ingredient $F_{\mu\nu}(A)$ by  $e_{[\mu\nu]}[B]$ and Lorentz covariant versions of $I_{ij} $ and $I_{j}$.

\section{Massive spin-1 models}

The simplest way to obtain the massive model from the massless one is to add the usual Proca mass term $-m^2A^2/2$ to our first order massless models, for instance, adding it to (\ref{lb}) and (\ref{lw}) and integrating over the vector field we obtain respectively, the massive Kalb-Ramond model \cite{kr} and the symmetric massive model of \cite{renato1},

\bea {\cal L}_{KR} &=& \frac 12 (\p^{\mu}B_{\mu\nu})^2 + \frac{m^2}4 B_{\mu\nu}B^{\mu\nu}  \quad , \label{kr} \\
{\cal L}_W^m &=& (\p^{\mu}W_{\mu\nu})^2 + \frac{m^2}2 \left( W_{\mu\nu}^2 - \frac{W^2}{D-1}\right) \quad . \label{lwm} \eea 

\no where we have slightly redefined the fields $(B_{\mu\nu},W_{\mu\nu})\to m (B_{\mu\nu},W_{\mu\nu}/\sqrt{2})$.

Similarly, adding the Proca mass term to  (\ref{sns}) and integrating over the vector field we obtain a massive non symmetric model. In order to investigate  the duality with the Proca theory in the presence of interactions we include matter fields and suggest the parent action:

\be\label{naaosimetricosemsimetria}
  S_P = \int\, d^Dx \Big\{ b\, e_{\mu \nu} e^{\mu \nu} + c\, e_{\mu \nu} e^{\nu \mu}  - \frac{(b+c)\, e^2}{D-1} + \partial_\mu A_\nu  e^{\mu \nu} - \frac{b\, m^2 A_\mu A^\mu}{4 (b^2-c^2)} + J_\mu A^\mu + e_{\mu \nu} J^{\mu \nu} + \mathcal{L}_M   \Big\} , \label{sp2}
\ee
with $J_\mu$ and $J_{\mu \nu}$ functions of the matter fields $\psi_l$ while $\mathcal{L}_M=\mathcal{L}_M[\psi_l]$ is the pure matter Lagrangian. The functional integral over $e_{\mu \nu}$ furnishes the Proca theory,

\bea\label{MPQUA}
  S^I_{MP} = \int\, d^Dx \Big\{ &-& \frac{bF_{\mu \nu} F^{\mu \nu}}{8(b^2-c^2)} - \frac{m^2 b\, A^\mu A_\mu}{4 (b^2-c^2) } + A_\mu J^\mu - \frac{J^{\mu \nu}}{4(b^2-c^2)} \left[ b J_{\mu \nu} -  c J_{\nu \mu} - (b-c) \eta_{\mu \nu} J \right] \nonumber\\
    &-& \frac{J^{\mu \nu}}{2(b^2-c^2)} \left[ b \partial_\mu A_\nu - c \partial_\nu A_\mu  - (b-c) \eta_{\mu \nu} \partial^\alpha A_\alpha \right] + \mathcal{L}_M \Big\}, \label{procai}
\eea

\no On the other hand, integrating $A_\mu$ leads to the new non symmetric massive model linearly coupled to sources:
\bea\label{modeloeI}
  S^I[e]=  \int\, d^Dx \Big\{\frac{(b^2-c^2)}{b\, m^2} \partial_\mu e^{\mu \nu} \partial^\alpha e_{\alpha \nu}  + b\, e_{\mu \nu} e^{\mu \nu} &+& c\, e_{\mu \nu} e^{\nu \mu}  - \frac{(b+c) e^2}{D-1} +     e_{\mu \nu} J^{\mu \nu} \nonumber\\
    &+& \frac{(b^2-c^2)}{b m^2} \left(   2 e^{\mu \nu} \partial_\mu J_\nu + J^\mu J_\mu\right) + \mathcal{L}_M  \Big\}. \nn\\ \label{nsi}
\eea

\no The canonical analysis of the new model, at free level, is 
carried out in the appendix. In both dual theories (\ref{procai}) and (\ref{nsi}), linear and quadratic (Thirring like) terms in matter currents are generated. First order functional derivatives of (\ref{MPQUA}) and (\ref{modeloeI}) with respect to the source $J_{\mu}$ , as if it was an external current, suggest  the following correspondence 
between the Proca and the non symmetric model,

\be\label{mappeA}
   A_\nu \longleftrightarrow -\frac{2 (b^2-c^2) }{b m^2} \left( \partial^\alpha e_{\alpha \nu} -J_\nu \right), 
\ee

\no Let us show that (\ref{mappeA}) is confirmed when we look at the equations of motion of $ S^I_{MP}$ and $ S^I[e]$, including the matter fields. First, the equations of motion $\delta\,S^I[e]/\delta\, e^{\mu\nu}=0$ are
\be\label{10099}
    b\, e_{\mu \nu} + c\,  e_{\nu \mu} - \frac{(b+c) e}{D-1} \eta_{\mu \nu} - \frac{(b^2 - c^2)}{b m^2} \partial_\mu \partial^\alpha e_{\alpha \nu} + \frac{J_{\mu \nu}}{2} + \frac{(b^2-c^2)}{b m^2} \partial_\mu J_\nu = 0,
\ee
and its  trace is 
\be
    -\frac{(b+c) e }{D-1} - \frac{(b^2-c^2)}{b m^2} \partial^2 e  + \frac{J}{2} + \frac{(b^2-c^2)}{b m^2} \partial^\alpha J_\alpha = 0.
\ee
Replacing the trace back in (\ref{10099}) we get
\bea
    E_{\mu \nu} \equiv  b\, e_{\mu \nu} + c\, e_{\nu \mu} &-& \left[ -
    \frac{(b^2-c^2)}{b m^2} \partial^2 e  + \frac{J}{2} + \frac{(b^2-c^2)}{b m^2} \partial^\alpha J_\alpha \right] \eta_{\mu \nu} - \frac{(b^2 - c^2)}{b m^2} \partial_\mu \partial^\alpha e_{\alpha \nu} \nonumber\\
    &+& \frac{J_{\mu \nu}}{2} + \frac{(b^2-c^2)}{b m^2} \partial_\mu J_\nu = 0.
\eea
From $b \, E_{\mu  \nu} - c\, E_{\nu \mu} = 0$ we have
\bea\label{10098}
     e_{\mu  \nu} &=&  - \frac{1}{2(b^2-c^2)} \left[ b\, J_{\mu  \nu} - c J_{\nu \mu} - (b-c) \eta_{\mu \nu} J \right]  \nonumber\\
    &+&  \frac{1}{b m^2} \left[ b \partial_\mu (\partial^\alpha e_{\alpha \nu} - J_\nu ) - c \partial_\nu (\partial^\alpha e_{\alpha \mu} - J_\mu ) - \eta_{\mu \nu} (b-c) ( \partial^2 e - \partial_\mu J^\mu ) \right],
\eea
the $\partial^\mu$ divergence gives
\be\label{eqeq}
     \frac{ \left( \square - m^2 \right)}{ m^2}  \left(\partial^\alpha e_{\alpha \nu}- J_\nu \right)  - \frac{\partial_\nu}{ m^2} \left(\partial^2 e - \partial^\mu J_\mu \right) 
    - \frac{1}{2 (b^2-c^2)} \left[ b \partial^\alpha J_{\alpha \nu} - c  \partial^\alpha J_{\nu \alpha} - (b-c) \partial_\nu J \right] \nonumber\\
    - J_\nu  = 0,
\ee

\no While the equations of motion $\delta S^I_{MP}/\delta A^{\nu} = 0$ are 
\be\label{eqeq2}
    \frac{b}{2(b^2-c^2)} \left( \square A_\nu - \partial_\nu \partial^\mu A_\mu \right) - \frac{m^2 b A_\nu }{2(b^2-c^2)} + J_\nu + \frac{1}{2(b^2-c^2)} \left[ b \partial^\alpha  J_{\alpha \nu} - c \partial^\alpha J_{\nu \alpha} - (b-c) \partial_\nu J \right] = 0.
\ee
By inspection we see that the map (\ref{mappeA}) guarantees the correspondence between (\ref{eqeq}) and (\ref{eqeq2}). Regarding the matter equations of motion,  using (\ref{10098}), $\delta_\psi S^I[e] = 0$ gives
\bea
    \frac{\delta \mathcal{L}_M }{\delta \psi_l} &-& \frac{2(b^2-c^2)}{b m^2} \left( \partial^\alpha e_{\alpha \nu} - J_\nu \right) \frac{\delta J^\nu}{\delta \psi_l} + \Bigg\{ - \frac{1}{2(b^2-c^2)} \left[ b J_{\mu  \nu} - c J_{\mu \nu} - (b-c) \eta_{\mu \nu} J \right] \nonumber\\
    &+& \frac{1}{b m^2} \left[ b \partial_\mu (\partial^\alpha e_{\alpha \nu} - J_\nu ) - c \partial_\nu (\partial^\alpha e_{\alpha \mu} - J_\mu ) - \eta_{\mu \nu} (b-c) ( \partial^2 e - \partial_\mu J^\mu ) \right] \Bigg\} \frac{\delta J^{\mu \nu}}{\delta \psi_l} = 0. \nonumber\\ \label{3212}
\eea
while $\delta_\psi S^I_{MP}=0$ furnishes
\be\label{eqeq3}
    \frac{\delta \mathcal{L}}{\delta \psi_l } - \Bigg\{  \frac{\left[ b J_{\mu \nu} - c J_{\nu \mu} - (b-c) J \eta_{\mu \nu} \right]}{2(b^2-c^2)}  
    + \frac{\left[ b \partial_\mu A_\nu  - c \partial_\nu A_\mu - (b-c) \eta_{\mu \nu} \partial^\mu A_\mu \right]}{2(b^2-c^2)}  \Bigg\}  \frac{\delta J^{\mu \nu}}{\delta \psi_l} \nonumber\\
    + A_\mu \frac{\delta J^\mu}{\delta \psi_l} =0.
\ee

Therefore, the matter field equations are also equivalent via the map (\ref{mappeA}). Thus, the master action (\ref{sp2}) shows the duality between the interacting theories (\ref{procai}) and (\ref{nsi}) which have non trivial Thirring like self-interacting terms. A similar duality involving the KR model is established in \cite{maluf}.

\section{Massless limit and dimensional reduction}

In the present section we investigate some subtleties of the massless limit of higher rank models and its relationship to dimensional reduction. 

The massless limit of the Proca model corresponds to the Maxwell theory which describes  the helicities $\pm 1$, so we only loose the
longitudinal mode (helicity $0$) as $m\to 0$. Such mode can be recovered if we  introduce a scalar Stueckelberg field in the usual way $A_{\mu} \to A_{\mu} + \p_{\mu}\phi /m$ before $m\to 0$.  In the case of the massive antisymmetric field (\ref{kr}) the discontinuity is more severe. We have a ``spin jumping'' \cite{dts} as $m\to 0$. Namely, the Lagrangian $(\p^{\mu}B_{\mu\nu})^2$ describes a massless spin zero field as one can check \cite{dw}
by going to the first order equivalent model
${\cal L}_C = -C^{\mu}C_{\mu} + 2\, C_{\mu} \p_{\alpha}B^{\alpha\mu}$ . Integrating on $B^{\alpha\mu}$ we have $\p_{\alpha}C_{\nu}- \p_{\nu}C_{\alpha} =0 $, thus leading to $ C_{\mu} = \p_{\mu} \phi $, so ${\cal L}_C $ becomes $- (\p_{\mu}\phi)^2$. The most singular massless limit occurs however, in the symmetric and non symmetric cases. After redefining $e_{\mu\nu} \to m \sqrt{b/(b^2-c^2)} \, e_{\mu\nu}$ and dropping the sources, the non symmetric massive spin-1 model (\ref{modeloeI}) can be written as

\be  {\cal L}_{NS} = \partial_\mu e^{\mu \nu} \partial^\alpha e_{\alpha \nu}  + \frac{b\, m^2}{b^2-c^2} \left( b \, e_{\mu \nu} e^{\mu \nu} + c \, e_{\mu \nu} e^{\nu \mu}  - \frac{(b+c) e^2}{D-1}\right) \quad . \label{lns} \ee

\no In the massless limit we have $( \partial_{\mu} e^{\mu \nu})^2$ which has no particle content as one can see by replacing $B_{\alpha\mu} \to e_{\alpha\mu}$ in ${\cal L}_C$ and integrating over $e_{\mu\nu}$ leading to $\p_{\mu}C_{\nu}=0$ which, assuming vanishing fields at infinity,  requires $C_{\mu} =0$. This is the same solution of the constraint $\p_{\mu}C_{\nu}+ \p_{\nu}C_{\mu} =0 $ appearing in the massless limit of the symmetric theory (\ref{lwm}).
Therefore, no propagating degree of freedom is left in the $m\to 0$ limit of (\ref{lwm}) and of (\ref{lns}) in arbitrary dimensions.

On the other hand, it is known that the Kaluza-Klein (KK) dimensional reduction  from $D+1$ to $D$ dimensions of the massless spin-1 Maxwell theory, when restricted to only one massive mode,  furnishes the corresponding massive spin-1 (Proca) model. Likewise, the KK reduction of the linearized massless spin-2 model, linearized Einstein-Hilbert about flat space, leads to the massive spin-2 Fierz-Pauli model, see for instance \cite{kmu}. This is also true for higher spins \cite{ady,rss,bhr}. So we learn that the KK dimensional reduction produces  the correct mass term without changing the spin of the particle in their lowest rank field representation. However, considering the previous remarks about the singular massless limit of (\ref{kr}),(\ref{lwm}) and (\ref{lns}), it is clear  that, differently from Proca, those massive models in $D$ dimensions can not stem from the KK dimensional reduction of their massless limits in $D+1$ dimensions. Henceforth we address the question whether those massive models with singular massless limits could also be obtained from a KK dimensional reduction of some massless model.

\subsection{Spin-0 Curtright and Freund model}

The problem of singular massless limit already appears in the higher rank description of massive spin-0 particles of \cite{curt}. We start with a first order description of massless scalar particles in $D$-dimensions,

\be S_{s=0}^{m=0}= \frac 12 \int \, d^D\, x \left(  A^{\mu}A_{\mu} + 2\,  A^{\mu}\p_{\mu}\phi \right) \quad , \label{soo} \ee

\no The integral over the vector field leads to the usual $\phi\Box\phi/2$ massless scalar particle. On the other hand, if we add the mass term $-m^2\phi^2/2$ and integrate over $\phi$ we obtain the vector description of massive  spinless particles given in \cite{curt}, after $A_{\mu}\to m A_{\mu}$,

\be S_{s=0}^m = \frac 12 \int \, d^D\, x \left\lbrack (\p^{\mu}A_{\mu})^2 + m^2\, A^{\mu}A_{\mu} \right\rbrack\quad , \label{som} \ee

\no The model (\ref{som}) is dual to the usual Klein-Gordon scalar theory at $m\ne 0$, however the massless limit of (\ref{som}) is singular since the first term $(\p^{\mu}A_{\mu})^2$ has no particle content. Thus, (\ref{som}) does not come from the dimensional reduction\footnote{In $(D+1)$ dimensions we use capital Latin letters to denote the $(D+1)$-dimensional indices, $(A,B,M,N,\ldots=0,1,2,\ldots,D)$
while in $D$ dimensions we use Greek letters $(\alpha,\beta,\mu,\nu,\ldots=0,1,2,\ldots,D-1)$.} of $(\p^{M}A_{M})^2$ with $M=0,1,\cdots,D$. However, integrating over $\phi$ in (\ref{soo}) leads to the constraint $\p^{\mu}A_{\mu}=0$ whose general solution  $A_{\nu}= \p^{\mu}B_{\mu\nu}$ plugged back in the action leads to the massless two-form field action for a spinless particle,  in
$D+1$ dimensions becomes

\be S_{s=0}^*= \frac 12 \int \, d^{D+1}\, x \, (\p^{M}B_{MN})^2\quad . \label{s02}  \ee

The action (\ref{s02}) is invariant under $\delta{B}_{MN}=\partial^{A}\Omega_{[AMN]}$. Now we proceed with the KK dimensional reduction of $S_{s=0}^*$.
Let us compact the last spatial dimension $x^{D}\equiv{y}$ in a circle of radius $R=1/m$
and keep only one massive mode. So the field and gauge parameter are redefined respectively as

\bea
B_{MN}(x^{\alpha},y)\rightarrow\left\{\begin{array}{l}
B_{\mu\nu}=\sqrt{\frac{m}{\pi}}\,B_{\mu\nu}(x)\cos{my}\\
B_{D\mu}=\sqrt{\frac{m}{\pi}}\,A_{\mu}(x)\sin{my}
\end{array}\right.\label{Reduc-B}\quad,
\eea
\bea
\Omega_{[AMN]}(x^{\alpha},y)\rightarrow\left\{\begin{array}{l}
\Omega_{[\alpha\mu\nu]}=\sqrt{\frac{m}{\pi}}\,\Omega_{[\alpha\mu\nu]}(x)\cos{my}\\
\Omega_{[D\mu\nu]}=\sqrt{\frac{m}{\pi}}\,\Theta_{[\mu\nu]}(x)\sin{my}
\end{array}\right. \label{param-B}\quad.
\eea

Substituting (\ref{Reduc-B}) in (\ref{s02}) and integrating over $y$
we obtain the following massive action in $D$ dimensions

\bea
S^{Stueck}_{s=0} = \frac{1}{2}\int \, d^{D}\, x \Bigg[(\partial^{\mu}A_{\mu})^2 + m^{2}\Big(A_{\mu}+\frac{\partial^{\nu}B_{\nu\mu}}{m}\Big)^{2}\Bigg]\quad.
\label{Stu-s02} \eea
It is the Stueckelberg version of (\ref{som}) where $B_{\mu\nu}$ acts as a Stueckelberg field after the dimensional reduction.
This action is invariant under the transformations
$\delta{A}_{\mu}=\partial^{\alpha}\Theta_{[\alpha\mu]}$ and $\delta{B}_{\mu\nu}=-m\Theta_{[\mu\nu]}+\partial^{\alpha}\Omega_{[\alpha\mu\nu]}$.
In the massless limit we obtain $\mathcal{L} \sim (\partial^{\mu}A_{\mu})^{2}+(\partial^{\mu}B_{\mu\nu})^{2}$,
where $(\partial^{\mu}A_{\mu})^{2}$ has not particle content and $(\partial^{\mu}B_{\mu\nu})^{2}$ describes a massless spin-0 particle.
Therefore the massless limit is smooth, preserving both the symmetry and the number of degrees of freedom. Since $B_{\mu\nu}$ is pure gauge in (\ref{Stu-s02}) we can fix $B_{\mu\nu}=0$ at action level \cite{moto} and claim that (\ref{som}) comes from the KK dimensional reduction of (\ref{s02}). 

\subsection{Spin-1 Kalb-Ramond model}
In the spin-1 case, the Kalb-Ramond model (\ref{kr})
cannot be obtained from the dimensional reduction of its massless limit $(\partial^{M}B_{MN})^{2}$,
since this term describes a massless spin-0 particle.
Let us then consider the Lagrangian density (\ref{ltt}) which describes a massless spin-1 particle
in terms of a totally antisymmetric tensor. In $D+1$ dimensions, the spin-1 analogue of (\ref{s02}) is given by

\bea
S^{\ast}_{TT} = \frac{1}{4}\int{d}^{D+1}x \,(\partial_{A}T^{[AMN]})^{2} \quad, \label{LTT}
\eea
which is invariant under $\delta{T}_{AMN}=\partial^{B}\Omega_{[BAMN]}$.
Performing the KK dimensional reduction, the field and gauge parameter are redefined analogously to (\ref{Reduc-B}-\ref{param-B})

\bea
T_{[AMN]}(x^{\alpha},y)\rightarrow\left\{\begin{array}{l}
T_{[\alpha\mu\nu]}=\sqrt{\frac{m}{\pi}}\,T_{[\alpha\mu\nu]}(x)\cos{my}\\
T_{[D\mu\nu]}=\sqrt{\frac{m}{\pi}}\,B_{\mu\nu}(x)\sin{my}
\end{array}\right.\label{Reduc-T}\quad,
\eea
\bea
\Omega_{[BAMN]}(x^{\alpha},y)\rightarrow\left\{\begin{array}{l}
\Omega_{[\beta\alpha\mu\nu]}=\sqrt{\frac{m}{\pi}}\,\Omega_{[\beta\alpha\mu\nu]}(x)\cos{my}\\
\Omega_{[D\alpha\mu\nu]}=\sqrt{\frac{m}{\pi}}\,\Theta_{[\alpha\mu\nu]}(x)\sin{my}
\end{array}\right. \label{param-T}\quad .
\eea

\no Back with (\ref{Reduc-T}) in (\ref{LTT}) we obtain (after an integration in $y$)
the Stueckelberg version of the Kalb-Ramond action (\ref{kr})

\bea
S^{Stueck}_{KR} = \frac{1}{2}\int \, d^{D}\, x \Big[(\partial^{\mu}B_{\mu\nu})^2 + \frac{m^{2}}{2}\Big(B_{\mu\nu}+\frac{\partial^{\alpha}T_{[\alpha\mu\nu]}}{m}\Big)^{2}\,\Big] \label{Skr} ,
\eea
which is invariant under the transformations $\delta{B}_{\mu\nu}=\partial^{\alpha}\Theta_{[\alpha\mu\nu]}$
and $\delta{T}_{[\alpha\mu\nu]}=-m\Theta_{[\alpha\mu\nu]}+\partial^{\beta}\Omega_{[\beta\alpha\mu\nu]}$.
Note that after reduction $T_{[\alpha\mu\nu]}$ act as a Stueckelberg field while $B_{\mu\nu}$ is the main field.
Taking $m\rightarrow0$ in (\ref{Skr}) we obtain
$\mathcal{L}\sim (\partial^{\mu}B_{\mu\nu})^{2}+\frac{1}{2}(\partial^{\alpha}T_{[\alpha\mu\nu]})^{2}$.
The term $(\partial^{\mu}B_{\mu\nu})^{2}$ describes a massless spin-0 particle while  $(\partial^{\alpha}T_{[\alpha\mu\nu]})^{2}$
describes a massless spin-1 particle.
 We have a smooth massless limit with $3$ degrees of freedom in $D=4$. Once again we can get rid of the pure gauge fields fixing $T_{[\alpha\mu\nu]}=0$ at action level and conclude that the massive Kalb-Ramond model (\ref{kr}) is the KK dimensional reduction of (\ref{LTT}).

\subsection{Spin-1 symmetric model}
Similar to the Kalb-Ramond model, the symmetric massive spin-1 model (\ref{lwm}) can not be obtained via
dimensional reduction of $(\partial^{M}W_{MN})^{2}$.
Let us now perform the dimensional reduction of the higher rank and higher order model $\mathcal{L}_{DTS}$ (\ref{ldts}),
which in $D+1$ is given by the action
\bea
S^{\ast}_{DTS} = \frac{1}{2}\int{d}^{D+1}x\,\Bigg[(\partial_{M}\partial_{N}B^{[MA][NB]})^{2}-\frac{(\partial_{M}\partial_{N}B^{MN})^{2}}{D}\Bigg]\quad,
\label{LBB}
\eea
where $B_{MN} = \eta^{AB} B_{[MA][NB]}$. This model is invariant under
\bea
\delta{B}_{[MA][NB]}&=&\partial^{L}\big(\Omega_{[LMA][NB]}+\Omega_{[LNB][MA]}\big)
+(\eta_{MN}\eta_{AB}-\eta_{MB}\eta_{AN})\varphi \quad .
\eea
\no Performing the dimensional reduction, the field $B_{[MA][NB]}$ is redefined as
\bea
B_{[MA][NB]}(x^{\alpha},y)\rightarrow\left\{\begin{array}{l}
B_{[\mu\alpha][\nu\beta]}=\sqrt{\frac{m}{\pi}}\,B_{[\mu\alpha][\nu\beta]}(x)\cos{my}\\
B_{[D\alpha][\nu\beta]}=\sqrt{\frac{m}{\pi}}\,Y_{\alpha[\nu\beta]}(x)\sin{my}\\
B_{[D\alpha][D\beta]}=\sqrt{\frac{m}{\pi}}\,W_{\alpha\beta}(x)\cos{my}
\end{array}\right.\label{Reduc-BB}\quad.
\eea
\no The fields $W_{\mu\nu}$, $Y_{\alpha[\nu\beta]}$ and $B_{[\mu\alpha][\nu\beta]}$
transform according to
\bea
\delta{W}_{\mu\nu} &=&
-\partial^{\sigma}\big(\Psi_{\mu[\sigma\nu]}+\Psi_{\nu[\sigma\mu]}\big)+\eta_{\mu\nu}\varphi \quad, \label{dw}\\
\delta{Y}_{\alpha[\nu\beta]} &=&
-\partial^{\sigma}\big(\Theta_{[\sigma\alpha][\nu\beta]}-\Pi_{\alpha[\sigma\nu\beta]}\big)-m\,\Psi_{\alpha[\nu\beta]} \quad, \label{dy}\\
\delta{B}_{[\mu\alpha][\nu\beta]} &=&
\partial^{\sigma}\big(\Omega_{[\sigma\mu\alpha][\nu\beta]}+\Omega_{[\mu\alpha][\sigma\nu\beta]}\big)+2m\Theta_{[\mu\alpha][\nu\beta]}
+\eta_{\mu\nu}\eta_{\alpha\beta}\varphi-\eta_{\mu\beta}\eta_{\alpha\nu}\varphi \quad \label{db},
\eea
where the gauge parameters
$\Psi_{\mu[\alpha\nu]}$, $\Theta_{[\sigma\alpha][\nu\beta]}$, $\Pi_{\alpha[\sigma\nu\beta]}$ and $\Omega_{[\lambda\mu\alpha][\nu\beta]}$
come from the dimensional reduction of the parameters $\Omega_{[LMA][NB]}(x)$. Substituting (\ref{Reduc-BB}) in (\ref{LBB}) and integrating  $y$ we find
\be
S^{S}_{W} =
\int{d}^{D}x\,\left\lbrace\Bigg[\frac{(D-1)}{2D}\left(\partial_{\mu}\partial_{\nu}\widetilde{W}^{\mu\nu}+\frac{m^2\,  \widetilde{W}}{D-1} \right)^2
+ \Bigg[m^{2}(\partial^{\mu}\widetilde{W}_{\mu\nu})^{2} +\frac{m^{4}}{2}\Big(\widetilde{W}^{\,2}_{\mu\nu}-\frac{\widetilde{W}^{\,2}}{(D-1)}\Big)\Bigg]  \right\rbrace , \label{S-WW}
\ee
where we have defined
\bea
\widetilde{W}_{\mu\nu}=W_{\mu\nu}-\frac{\partial^{\alpha}(Y_{\mu[\alpha\nu]}+Y_{\nu[\alpha\mu]})}{m}
-\frac{\partial^{\alpha}\partial^{\beta}B_{[\mu\alpha][\nu\beta]}}{m^{2}} \quad ,
\eea
and $\widetilde{W}=\eta^{\mu\nu}\widetilde{W}_{\mu\nu}$.
The fields $Y_{\mu[\alpha\nu]}$ and $B_{[\mu\alpha][\nu\beta]}$ act as Stueckelberg fields, see (\ref{dy}) and (\ref{db}),
whereas $\widetilde{W}_{\mu\nu}$ becomes the physical field. 
The action (\ref{S-WW}) is invariant under the transformation
\be
\delta_{\varphi}\widetilde{W}_{\mu\nu} = \frac{1}{m^{2}}[\partial_{\mu}\partial_{\nu}\varphi-(\square-m^{2})\eta_{\mu\nu}\varphi] \quad.
\ee
\no The factor 
$f\equiv\partial_{\mu}\partial_{\nu}\widetilde{W}^{\mu\nu}+m^{2}\widetilde{W}/(D-1)$ which appears squared in (\ref{S-WW}) is pure gauge: $\delta_{\varphi} f = -m^2D\, \varphi/(D-1)$. So we can introduce a pure gauge scalar field $\phi$  to lower the order of  (\ref{S-WW}) by replacing the square term by $[(D-1)/D](-\phi^2/2 + \phi \, f)$. We can rewrite the final Lagrangian in the form of a Stueckelberg version of  the symmetric model (\ref{lwm}) up to an overall $m^2 $ factor,
\bea
S_{W} &=& m^2 \, 
\int{d}^{D}x\,\Bigg[(\partial^{\mu}\overline{W}_{\mu\nu})^{2} +\frac{m^{2}}{2}\Big(\overline{W}^{\,2}_{\mu\nu}-\frac{\overline{W}^{\,2}}{(D-1)}\Big)\Bigg] \quad. \label{S-WW2}
\eea
\no where the field $\overline{W}_{\mu\nu}$ defined below is gauge invariant since $\delta\phi = - D\, m^2 \varphi/(D-1)$ and
\be \widetilde{W}_{\mu\nu} = \overline{W}_{\mu\nu} - \frac{D}{m^4(D-1)}\left\lbrack  \p_{\mu}\p_{\nu}\, \phi - (\Box - m^2)\, \eta_{\mu\nu}\phi \right\rbrack  \quad . \label{wbar} \ee
\no The Stueckelberg fields in (\ref{S-WW2}) provide a smooth massless limit once again. Taking $m\rightarrow0$ in (\ref{S-WW2}) we obtain
$\mathcal{L}=(\partial_{\mu}\partial_{\alpha}Y^{\mu[\alpha\nu]})^{2} + \mathcal{L}_{DTS}(B)$.
The model $\mathcal{L}_{DTS}(B)$ describes a massless spin-1 particle in $D$ dimensions
while $(\partial_{\mu}\partial_{\alpha}Y^{\mu[\alpha\nu]})^{2}$ describes a massless spin-0 particle,
as will be shown below.
The second-order version of $\mathcal{L}_{YY}=(\partial_{\mu}\partial_{\alpha}Y^{\mu[\alpha\nu]})^{2}$ is
\bea \mathcal{L}^{(2)} = -A_{\mu}A^{\mu}+Y^{\mu[\alpha\nu]}\partial_{\mu}(\partial_{\alpha}A_{\nu}-\partial_{\nu}A_{\alpha}) \quad,
\eea
where $A_{\mu}$ is an auxiliary vector field.
Integrating  $A_{\mu}$ we recovery $\mathcal{L}_{YY}$.
On the other hand, integrating  $Y^{\mu[\alpha\nu]}$ we obtain the constraint
$\partial^{\mu}(\partial^{\alpha}A^{\nu}-\partial^{\nu}A^{\alpha})=0$
whose general solution is $A^{\nu}=\partial^{\nu}\phi$. Back in $\mathcal{L}^{(2)}$
we obtain $\mathcal{L}(\phi)=\phi\square\phi$.
Therefore, the total number of degrees of freedom  of the symmetric massive model, i.e. $D-1$, is preserved in the massless limit. Since the fields $Y_{\mu[\alpha\nu]}$ a, $B_{[\mu\alpha][\nu\beta]}$ and  $\phi$ are pure gauge Stueckelberg fields they can all be set to zero  at action level and  we recover the symmetric model (\ref{lwm}).

\section{Spin-1 non symmetric model}

Starting with the $D+1$ massless theory
\bea
S_{BB} = \int d^{D+1} x \Bigg[ b\, \partial^M B_{[MN] A} \partial_K B^{[KN]A} + c \, \partial^M B_{[MN] A} \partial_K B^{[KA]N} - \frac{(b+c)}{D} \left( \partial^M B_M \right)^2 \Bigg] ,
\eea
where $B_{M}=\eta^{AB}B_{[MA]B}$. The action $S_{BB}$ is invariant by 
\bea
\delta B_{[MN]A} = \left( \eta_{AM} \partial_N - \eta_{AN} \partial_M \right) \Lambda + \Omega^T_{[MN]A},
\eea
we decompose the fields according to:
\bea
B_{[MA]N}(x^{\alpha},y)\rightarrow\left\{\begin{array}{l}
B_{[\mu\alpha]\nu}=\sqrt{\frac{m}{\pi}}\,B_{[\mu\alpha]\nu}(x)\cos{my}\\
B_{[D\alpha]\nu}=\sqrt{\frac{m}{\pi}}\,e_{\alpha\nu}(x)\sin{my}\\
B_{[D\alpha]D}=\sqrt{\frac{m}{\pi}}\,A_{\alpha}(x)\cos{my}\\
B_{[\alpha \mu]D} = \sqrt{\frac{m}{\pi}}\,V_{[\alpha\mu]}\sin{my}
\end{array}\right.
\eea
Performing the dimensional reduction gives 
\bea \label{S_BBD}
S^{stueck}_{BB} = \int d^D x  \Bigg[ b \left( \partial_\mu \tilde{e}^{\mu \nu} \right)^2 + b m^2 \tilde{e}^2_{\mu \nu} + c m^2 \tilde{e}_{\mu \nu} \tilde{e}^{\nu \mu} + m^2 \tilde{A}^2_\nu + 2 c m  \partial_\mu \tilde{e}^{\mu \nu} \tilde{A}_{\nu} + (b+c) \left( \partial^\mu \tilde{A}_\mu \right)^2\nonumber\\
- \frac{(b+c)}{D} \left(m \tilde{e} - \partial^\mu \tilde{A}_\mu \right)^2 \Bigg] , 
\eea
where 
\bea 
\tilde{e}_{\mu \nu} \equiv e_{\mu \nu} + \frac{\partial^\alpha B_{[\alpha \mu ] \nu}}{m} , \qquad \tilde{A}_\nu \equiv A_\nu - \frac{\partial^\alpha V_{[\alpha \nu ]} }{m} .
\eea
The above action is invariant by the gauge transformations 
\bea
\delta A_\mu = \partial_\mu \Lambda + \frac{\partial^\alpha \Theta_{[\alpha \mu]}}{m}, \qquad \delta e_{\mu \nu} = m \,\eta_{\mu \nu} \Lambda - \frac{\partial^\alpha \Omega_{[\alpha \mu ] \nu}}{m}, \\
\delta V_{[\mu \nu]} = \Theta_{[\mu \nu]} ,  \qquad \delta B_{[\mu \nu] \alpha} = \left( \eta_{\alpha \mu} \partial_\nu - \eta_{\alpha \nu} \partial_\mu \right) \Lambda + \Omega_{[\mu \nu] \alpha}.
\eea 
In the massless limit we have 
\bea
S^{stueck}_{BB} = \int d^D x \Bigg[ b \left(\partial_\mu e^{\mu \nu} \right)^2 +  (b+c) \left(\partial_\mu A^\mu \right)^2 + b \left( \partial^\mu B_{[\mu \nu] \alpha} \right)^2  + b \left( \partial^\mu V_{[\mu \nu]} \right)^2 \nonumber\\  + c\, \partial^\mu B_{[\mu \nu] \alpha} \partial_\beta B^{[\beta \alpha ] \nu}
- 2c \partial^\mu e_{\mu \nu} \partial_\alpha V^{[\alpha \nu]}  - \frac{(b+c)}{D} \left( \partial^\gamma B_\gamma - \partial^\mu A_\mu \right)^2 \Bigg].
\eea
If we redefine $(A_{\mu},e_{\alpha\mu})\to (\bar{\bar{A}}_{\mu}- B_{\mu}/(D-1),\bar{\bar{e}}_{\alpha\mu}+ c\, V_{[\alpha\mu]}/b)$ we decouple $\bar{\bar{A}}_{\mu}$ and $\bar{\bar{e}}_{\alpha\mu}$. Neglecting $(\p^{\mu}\bar{\bar{A}}_{\mu})^2$ and $(\p^{\alpha}\bar{\bar{e}}_{\alpha\mu})^2$ which have no particle content
 it follows that  
\bea
S^{stueck}_{BB} = \int d^D x \Bigg[ b \left(\partial^\mu B_{[\mu \nu] \alpha} \right)^2 +  c\, \partial^\mu B_{[\mu \nu] \alpha} \partial_\beta B^{[\beta \alpha ] \nu}   - \frac{(b+c)}{D-1} \left( \partial^\gamma B_\gamma \right)^2 + \frac{(b^2-c^2)}{b} \left(\partial^\mu V_{[\mu \nu]} \right)^2 \Bigg]
\eea
Displaying  a smooth massless limit. 

On the other hand at finite mass 
we can rewrite (\ref{S_BBD}) as 
\bea
S^{stueck}_{BB} = \int d^D x \Bigg[ b \left( \partial^\mu \tilde{e}_{\mu \nu} \right)^2 + m^2 \left( b \tilde{e}^2_{\mu \nu} + c \tilde{e}_{\mu \nu} \tilde{e}^{\nu \mu} \right)  + b m^2 \tilde{A}^2_\nu + 2 m c \tilde{A}^\nu \partial^\mu \tilde{e}_{\mu \nu} - \frac{(b+c)}{D-1}m^2 \tilde{e}^2\nonumber\\
+ \frac{m^2(b+c)}{D (D-1)} \left( \tilde{e} + \frac{D-1}m\partial^\mu \tilde{A}_\mu  \right)^2 \Bigg].\label{sbb2}
\eea
\no Similarly to what happened in the symmetric case of last subsection, we can replace the  square term in (\ref{sbb2}) by two terms involving a scalar field,  

\be \left( \tilde{e} + \frac{D-1}m\partial^\mu \tilde{A}_\mu  \right)^2 \leftrightarrow - \phi^2 + 2\, \phi \left\lbrack \tilde{e} + \frac{D-1}m\partial^\mu \tilde{A}_\mu \right\rbrack \ee

\no After such replacement and the Gaussian integration on the vector field we return to the non symmetric model
(\ref{lns})

\bea
S_{BB} = \int d^D x \Bigg[ \frac{(b^2-c^2)}{b} \left( \partial^\mu \bar{e}_{\mu \nu} \right)^2 + m^2 \left( b \bar{e}^2_{\mu \nu} + c \bar{e}_{\mu \nu} \bar{e}^{\mu \nu} \right) - \frac{(b+c)}{D-1}m^2 \bar{e}^2  \Bigg].
\eea 

\no where (use $\delta\phi =D\, m  \Lambda$) we have the gauge invariant combination

\be \bar{e}_{\mu \nu} \equiv \tilde{e}_{\mu\nu} + \frac{(\Box\theta_{\mu\nu}-m^2\eta_{\mu\nu})\phi}{D\, m^2} \quad . \label{bare} \ee

In summary, higher rank spin-1 massive models in $D$-dimensions can not be directly derived via KK dimensional reduction of their massless limit in $D+1$ but they stem from the reduction of higher rank massless dual models where the role of Stueckelberg field and physical field is interchanged.

\section{Conclusion}

In section 2 we have started from  a rather general first order version of the Maxwell theory in arbitrary dimensions with a rank-2 auxiliary field, see (\ref{S1}),  and derived in a unified way three higher rank descriptions of the photon dynamics: (\ref{ltt}),(\ref{ldts}) and (\ref{lbb}). They correspond respectively to antisymmetric ($B_{\mu\nu}$), symmetric ($W_{\mu\nu}$) and non symmetric ($e_{\mu\nu}$) auxiliary fields. The last case is a new one. Its equations of motion can be written as  zero curvature conditions (\ref{eombar}) and the Maxwell equations now follow from a trivial identity (\ref{id}). Although the number of independent  gauge invariants is much higher than in the usual vector description of the  Maxwell theory, we have shown that their correlation functions map into correlations functions  of the magnetic and electric fields and their derivatives in the usual  Maxwell theory,  see (\ref{map1}), (\ref{map2}) and (\ref{map3}). Our calculations  may be specially useful for the comparison of dual massless theories where the gauge symmetry makes the derivation of dual maps more subtle. Our approach is based on a non covariant solution of the constraint on the sources required by gauge invariance.
In section 3 we have investigated the duality between the massive Proca theory (\ref{procai}) and a new  non symmetric massive spin-1 model (\ref{nsi}), both coupled to matter fields. The duality requires the addition of quadratic terms in the matter currents (Thirring like terms). With the help of the dual map
(\ref{mappeA}) we have shown the equivalence of the equations of motion including matter fields as in the analogous Kalb-Ramond case \cite{maluf}. We point out that although any general rank-2 tensor can be decomposed into symmetric and antisymmetric components:
$e_{\mu\nu} = e_{(\mu\nu)} + e_{[\mu\nu ]}$, it turns out that the non symmetric model can not be reduced to the previous antisymmetric and symmetric models.

The higher rank massive models (\ref{kr}),(\ref{lwm}) and (\ref{lns}) have a singular massless limit. In the case of (\ref{kr}) the spin jumps from 1 to 0 as $m\to 0$ while for  (\ref{lwm}) and (\ref{lns}) the particle content becomes empty. This point is relevant for the recent discussion, see \cite{hell}, about a possible non equivalence between the Proca and the Kalb-Ramond models  after a given energy scale, intimately connected with the massless limit, is introduced. In section 4 we show that although those massive models  have a singular massless limit they
can still be obtained from higher rank massless models via the  usual KK dimensional reduction where the role of the main physical field and the Stueckelberg field is interchanged.

We are currently investigating the coupling of the massless model 
(\ref{lbb}) and of the massive non symmetric model (\ref{lns}) to gravity. An interesting issue is the behaviour of the massless theory under rigid conformal transformations, i.e., $g_{\mu\nu} \to e^{\lambda} g_{\mu\nu}$  with constant $\lambda$. In the Maxwell theory the metric factors $\sqrt{-g}\, g^{\mu\alpha}g^{\nu\beta} F_{\mu\nu}F_{\alpha\beta} $ lead to conformal invariance (traceless energy-momentum tensor) precisely in $D=4$. In the ${\cal L}_{BB}$ model we typically  have $\sqrt{-g}\, g^{\theta\alpha}g^{\eta\beta}g^{\gamma\mu}g^{\rho\nu} \nabla_{\theta}B_{[\alpha\mu]\nu}\nabla_{\eta}B_{[\beta\gamma]\rho}$. Using the invariance of the Christoffel symbol,  $\delta_{\lambda}\Gamma^{\rho}_{\mu\nu}=0$,  we see that we  only have conformal invariance now in $D=8$. Similarly, the higher order ${\cal L}_{DTS}$ model in (\ref{ldts}) is  conformal invariant only in $D=12$.  This might indicate the non equivalence of their gravitational effective action and consequently non equivalence of the dual models at quantum level. However, in the spin-0 case similar differences show up when we compare the usual scalar field $\sqrt{-g}g^{\mu\nu}\p_{\mu}\varphi \p_{\nu}\varphi $ with the massless two-form formulation $\sqrt{-g}g^{\mu\nu}g^{\gamma\alpha}g^{\beta\rho} \nabla_{\nu}B_{\mu\beta}\nabla_{\alpha}B_{\gamma\rho}$ of spin-0 particles. They are conformal invariant in $D=2$ and $D=6$ respectively, but it turns out that their effective actions differ by a topological term in $D=4$, see \cite{19,28,29,24}, and also  in arbitrary dimensions \cite{24}. Consequently their quantum energy momentum tensor coincide. It might be interesting to examine that point for the dual massless spin-1 model (\ref{lbb}).
Regarding the case of massive p-forms/(D-p-1)-forms duality, similar results (quantum equivalence) apply in $D=4$ \cite{58} and also in higher dimensions \cite{kt}, see also discussions in \cite{felipe1} for an interaction of the massive two-form with a scalar field. It is interesting to investigate the coupling to gravity also in the case of the non symmetric model (\ref{lns}).  

Last, as already mentioned in the introduction, we recall that the non symmetric model
(\ref{lns})  might be relevant for the search of Galileon-like Lagrangians for massive spin-1 particles \cite{gpt}-\cite{mayprd20}. The point is that Galileon-like terms can be systematically found from contractions with the Levi-Civita tensor and a general non symmetric tensor may allow new possibilities. Already at cubic level the symmetric components of the rank-2 tensor allow the definition of local super-renormalizable vertices for spin-1 particles in $D=4$ which is not possible\footnote{Notice that $B_{\mu\nu}B^{\mu\alpha}B_{\alpha\,\,\,\,}^{\!\!\quad\nu}$ vanishes identically and the cubic vertex $A^{\mu}A_{\mu}\p^{\nu}A_{\nu}$ is power counting renormalizable, not super-renormalizable.} for antisymmetric or vector   fields. Namely, $ {\cal L} = g_1 e^3 + g_2 e\, e_{(\mu\nu)}^2 + g_3  e\, e_{[\mu\nu]}^2 + g_4 \,  e_{(\mu\nu)}e^{[\mu\alpha]}e_{[\alpha \,\,\,}^{\!\!\quad\nu ]} + g_5 \,  e_{(\mu\nu)}e^{(\mu\alpha)}e_{(\alpha }^{\!\!\quad\nu )} $.

 We are currently investigating Galileon-like terms and its coupling to gravity in order to explore possible astrophysical and cosmological consequences  to be compared with the vector case (vector tensor theories) examined in \cite{gpt,fhkmt,jhknt,sky,bh1,bh2}.

\section{Acknowledgements}
The work of D.D. is partially supported by CNPq  (grant 313559/2021-0). The work of  F.A. da S.B is supported by CAPES (Brazil) finance code 001.

\section{Appendix} 

Here we analyse the canonical structure of the non symmetric model
(\ref{lns}) and show its equivalence to the Proca theory in the reduced phase space. First, it  is convenient to implement the invertible rotation
$e_{\mu\nu} \to (b\, e_{\mu\nu} - c\, e_{\nu\mu})/(b^2-c^2)$ and work with
\bea
S_{NS} = \int d^D x \Bigg[  \frac{\left( b e_{\mu \nu} e^{\mu \nu} - c e_{\mu \nu} e^{\nu \mu } \right)}{(b^2-c^2)} + \frac{d \, e^2}{(b+c)^2} + \frac{\left( b \partial^\alpha e_{\alpha \nu} - c \partial^{\alpha} e_{\nu \alpha} \right)^2 }{b m^2 (b^2-c^2)} \Bigg] ,
\eea
with the conjugate momenta
\bea
\Pi^{00} &=& - 2 \frac{\left( b \partial_\beta e^{\beta 0 } - c \partial_\beta e^{0 \beta}\right) }{b m^2 (b+c) } \qquad \Pi^{0l} = - 2 \frac{\left(b \partial_\beta e^{\beta l}- c \partial_\beta e^{l \beta} \right)  }{m^2 (b^2-c^2) } \\
\Pi^{l 0 } &=&  \frac{2c}{b} \frac{\left(b \partial_\beta e^{\beta l}- c \partial_\beta e^{l \beta} \right)  }{m^2 (b^2-c^2) } \qquad \Pi^{ij} = 0.
\eea
In addition to $\Pi^{ij}=0$ we also have the constraint 
\bea
c \, \Pi^{0l} + b \, \Pi^{l0} = 0.
\eea
The full Hamiltonian is 
\bea
  H = \int d^{D-1} x \Bigg[ - \frac{b m^2 (b+c)}{4(b-c)} (\Pi^{00})^2 + \frac{m^2 (b^2-c^2)}{4b} \Pi^{0l} \Pi^{0l} - \frac{\left( b \partial_l e^{l0} - c \partial_l e^{0l} \right)  }{b-c} \Pi^{00} - \frac{d e^2}{(b+c)} \nonumber\\
    \frac{\left(b \partial_k e^{kl} - c \partial_k e^{lk} \right)}{b} \Pi^{0l} - \frac{\left( b e_{\mu \nu} e^{\mu \nu} - ce_{\mu \nu} e^{\nu \mu} \right)    }{b^2-c^2} + \lambda_{ij} \Pi^{ij} + \lambda_l \left(b \Pi^{l0} + c \Pi^{0l} \right) \Bigg].
\eea
The consistency condition of $\frac{d}{dt}\left(b\Pi^{l0}+ c\Pi^{0l}\right)=0$ gives the secondary constraint
\bea
 \Gamma^l \equiv (b+c) \,\partial^l \Pi^{00}+ 2\, e^{l0} = 0,
 \eea
whose consistency condition fixes $\lambda_l$. The condition $\dot{\Pi}^{ij} = 0$ gives 
\bea
 \chi^{ij} \equiv \{ \Pi^{ij} \, , \, H \}  = \frac{ b \partial^i \Pi^{0j} - c \partial^j \Pi^{0i}  }{b} + 2 \frac{b e^{ij} - c e^{ji}  }{b^2-c^2} + 2 \frac{d \delta^{ij} e }{(b+c)^2} = 0,
\eea
and $\dot{\chi}^{ij}=0$ leads to
\bea
 \dot{\chi}^{ij} = \{ \chi^{ij} \, , \, H \} = - \frac{c \partial^j \partial^i \Pi^{00}}{b} + 2 \frac{b \partial^i e^{0j} - c\partial^j e^{0i}  }{b^2-c^2} - 2 \frac{c \left(b \partial^i e^{j0} - c \partial^{j} e^{i0} \right)  }{b (b^2-c^2)} \nonumber\\
    2 \frac{(b \lambda^{ij} - c \lambda^{ji} )}{b^2-c^2} + \frac{d b m^2\delta^{ij} \Pi^{00} }{b^2-c^2} + 2 \frac{d \delta^{ij} \left(b \partial_l e^{l0} - c\partial_l e^{0l} \right) }{(b+c) ((b^2-c^2)} +  2 \frac{d  \delta^{ij} \delta^{lm} \lambda_{lm} }{(b+c)^2} = 0.
\eea
The previous expression fixes the traceless part of  $\lambda_{ij}$ since $\delta_{ij} \chi^{ij}=0$ does not depend upon $\lambda^k_k$, but it is a new constraint:
\bea
 \phi \equiv \delta^{ij} \dot{\chi}^{ij}  = \left( \nabla^2 - m^2 \right) \Pi^{00} + 2\frac{ \partial_l e^{0l} }{(b+c)} = 0.
\eea
The requirement $\dot{\phi} = 0$ gives $e=0$ whose consistency condition fixes $\lambda^k_k$ ending the algorithm.  
The reduced Hamiltonian reads 
\bea
 H^R = \int d^{D-1} x \Bigg[  \frac{b \Pi^{00} \partial_l e_{0l} }{2(b-c)} + \frac{(b^2-c^2) m^2 \Pi^{0l} \Pi^{0l} }{4b} + \frac{(b^2-c^2) \tilde{F}^2_{lk} }{8b} + \frac{b e_{0l} e_{0l}}{b^2-c^2} \Bigg],
\eea
where $\tilde{F}_{lk} \equiv \partial_l \Pi^{0k} - \partial_k \Pi^{0l}$. In order to see that $H_R$ is positive defined we note from $\phi=0$ that 
\bea
\Pi^{00} = \frac{2 \partial_l e_{0l} }{(b+c) \left(\nabla^2-m^2 \right) }, 
\eea
so that the reduced Hamiltonian in terms of the independent quantities $(\Pi^{0l},e_{0l})$ reads 
\bea
 H^R = \int d^{D-1} x \Bigg[ \frac{(b^2-c^2) m^2 \Pi^{0l} \Pi^{0l} }{4b} + \frac{(b^2-c^2) \tilde{F}^2_{lk} }{8b} + \frac{b}{b^2-c^2} \left(  e^2_{0l} + \frac{(\partial_l e_{0l})^2 }{\nabla^2-m^2} \right) \Bigg], \label{rh1}
\eea
We can decompose $e_{0l}$ into 
\bea
e_{0l} = \partial_l \Phi + v^t_l,
\eea
where $\partial^l v^t_l=0$. Now it is clear that the last two terms in $H_R$ are non negative
\bea
\frac{b}{b^2-c^2} \int d^{D-1} x \Bigg[ v^t_l v^t_l + \frac{m^2 \phi \left( - \nabla^2 \right) \phi }{m^2 - \nabla^2} \Bigg] \ge 0 \quad .
\eea

Moreover, in order to show the canonical equivalence in the reduced phase space with the Proca theory we look at the Dirac-brackets. It is convenient to use the equivalent set of constraints:
\bea
    \chi_1^{ij} \equiv e^{ij} + \frac{(b^2-c^2) \partial^i \Pi^{0j} }{2b } = 0 , \qquad \chi_2^{ij} \equiv \Pi^{ij} = 0, \qquad\chi_3^l \equiv b \Pi^{l0} + c \Pi^{0l} = 0,\\
    \chi_4^l \equiv e^{l0} + \frac{(b+c) \partial^l \Pi^{00} }{2} = 0, \qquad
    \chi_5 \equiv \left( \nabla^2 - m^2 \right) \Pi^{00} + 2\frac{ \partial_l e^{0l}
    }{(b+c)} = 0, \\
    \chi_6 \equiv e^{00} + \frac{(b^2-c^2) \partial_l \Pi^{0l} }{2b} = 0.
\eea
Given the constraint matrix 
\bea
    C_{N M} \equiv \{ \chi_N \, , \, \chi_M \},
\eea
one can show that the matrix elements of the inverse are 
\bea
   &\left( C^{-1} \right)_{\chi^{lk}_2(x)  ,  \chi^{ij}_1(y) }& = \delta^l_i \delta^k_j \delta^{D-1} (x - y), \quad \left( C^{-1} \right)_{\chi^{m}_3(x)  ,  \chi^{ij}_2(y) } =  - \frac{(b^2-c^2)}{2bm^2} \partial^m_{x} \partial^i_{x} \partial^j_{x} \delta^{D-1} (x - y ),\nn\\
    &\left( C^{-1} \right)_{\chi_6(x) \, , \, \chi^{ij}_2 (y)}&  =   \frac{(b-c)}{bm^2}  \partial^i_{x} \partial^j_{x} \delta^{D-1} (x - y ), \qquad  \left( C^{-1} \right)_{\chi_6(x) \, , \, \chi_5 (y)} = \frac{\delta^{D-1}(x-y)}{m^2}, \nn\\
    &\left( C^{-1} \right)_{\chi_5(x) \, , \, \chi^{p}_3 (y)}& = - \frac{(b+c) \partial^p_{x} \delta^{D-1}(x-y) }{2m^2 b }, \qquad  \left( C^{-1} \right)_{\chi_6(x) \, , \, \chi^{p}_4 (y)} = \frac{2c \partial^p_{y} \delta^{D-1}(x- y) }{b (b+c) m^2},  \nn\\
    &\left( C^{-1} \right)_{\chi^l_4(x) \, , \, \chi^{p}_3 (y)}& = \frac{\delta^l_p \delta^{D-1}(x-y)}{b} - \frac{c \partial^l_{x} \partial^p_{x} \delta^{D-1}(x - y) }{b^2 m^2}, \nn
\eea
The other matrix elements are either zero or fixed by antisymmetry of the inverse matrix.  Now one can show that
\bea
     \{ e_{0l} (x,t) \, , \, e_{0k}(y,t) \}_D &=&
   \{ \Pi^{0l} (x,t) \, , \, \Pi^{0k}(y,t) \}_D = 0,\\
   \{ e_{0l} (x,t) \, , \, \Pi^{0k}(y,t) \}_D  &=& \left( \delta^k_l + \frac{\partial^l_{x} \partial^k_{y} }{m^2} \right) \delta^3 (x - y).
\eea
The previous brackets altogether with the reduced Hamiltonian can be simplified and brought to the Proca form. The inspiration comes from the equations of motion. The Hamiltonian equations of motion in addition to the constraints of the theory lead to
\bea
   m^2 e_{\mu \nu} =  \partial_\mu \partial^\alpha e_{\alpha \nu }, \qquad \partial^\nu e_{\alpha \nu} = 0.
\eea
They are equivalent to the Proca equations:  $(\square - m^2 ) A_\nu = 0$ and $\partial^\nu A_\nu = 0$ with the identification
\bea
    e_{\mu \nu} = \sqrt{\frac{b^2-c^2}{2 b}} \partial_\mu A_\nu
\eea
where the square root factor is chosen for future convenience. The above identification and the constraints $\chi^l_4 = 0 $ and $\chi_5 = 0$ inspires us the following change of coordinates in the phase space
\bea
    \Pi^{0l} &=& = - \sqrt{\frac{2b }{b^2-c^2}} A^l, \\
     e_{0l} &=& \sqrt{\frac{b^2-c^2}{2b}} \left( \pi^l - \frac{\partial^l \nabla \cdot \pi}{m^2} \right).
\eea
Indeed, using the Dirac-brackets we can show that 
\bea
    \{ A_l(x,t) \, , \, \pi^k(y,t) \}_D &=&  \, \delta^{D-1}(x- y ),\\
    \{ A_l(x,t) \, , \, A_k(y,t) \}_D &=& \{ \pi^l(x,t) \, , \, \pi^k(y,t) \}_D = 0.
\eea
In terms of the variables, $A_l$ and $\pi^l$ the reduced Hamiltonian (\ref{rh1}) reads
\bea
    H^R = \int \frac{d^{D-1} x}{2}  \left[ m^2 A^2  + \frac{(\nabla \cdot \pi )^2 }{m^2} + \pi^2 + \left( \nabla \times A \right)^2 \right].
\eea
This is just the reduced Proca Hamiltonian.

\end{document}